\documentclass{jps-cp}
\usepackage{txfonts} 
\usepackage{cite}
\usepackage{amsmath,amssymb,amsfonts}
\usepackage{amsmath}
\usepackage{algorithmic}
\usepackage{graphicx}
\usepackage{textcomp}
\usepackage{xcolor}
\usepackage{mathtools}
\usepackage{multirow}
\usepackage{booktabs}
\usepackage{enumerate}
\usepackage{subfigure}
\usepackage{multirow}
\usepackage{enumitem}
\usepackage{comment}

\title{Hodge Decomposition of the Remittance Network on the XRP ledger in the Price Hike of January 2018}
 
\author{
Yuichi \textsc{Ikeda}$^{1}$, 
Abhijit \textsc{Chakraborty}$^{1, 2}$
}
 
\inst{$^{1}$Graduate School of Advanced Integrated Studies in Human Survivability, Kyoto University, Kyoto 606-8306, Japan\\
$^{2}$RIKEN Interdisciplinary Theoretical and Mathematical Sciences Program, Saitama, 351-0198, Japan
}

\email{ikeda.yuichi.2w@kyoto-u.ac.jp}

\abst{
This study analyzes the remittance transaction recorded on the XRP ledger for ETH and USD from July 2017 to Jun 2018, including the bubble period in early 2018. 
Using the Hodge decomposition, we estimate the ``loop flow'' in the international remittance of cryptoassets during the bubble period.
We found characteristic differences between those fiat currencies and cryptoassets during the bubble period. For ETH, there was a significant increase in the loop flow during the cryptoasset price peak. This might be related to money laundering or arbitrage transaction.
There was a slight increase in the loop flow for USD during the cryptoasset price peak.
}

\kword{Ethereum, XRP ledger, remittance, Hodge decomposition, loop flow, anomaly detection}

\begin{document}
\maketitle

\section{Introduction}

The problem of international remittances, including those of immigrants and developing countries, has been attracting particular attention in recent years.
International remittances from migrants and others to their home countries are the most important source of funds for developing countries, and the demand for international remittances is growing as economic globalization progresses. 
In 2018, the global value of international remittances reached \$529 billion; except for China, the value of international remittances exceeded the sum of official development assistance, ODA, and foreign direct investment, FDI \cite{Barne2019}. 
Unlike ODA, which may not reach its intended recipients due to corruption, and FDI, which may not benefit as many people other than the invested industry sectors, international remittances can be sent directly to households in need of funds. For these reasons, international remittances have emerged as a viable means of raising funds for education, health, employment, and entrepreneurship. International remittances are also a consistent source of income for many households and are considered counter-cyclical because they are directed to households in the destination country that require more money during periods of low economic activity \cite{Frankel2011}.

Blockchain can be a fundamental technology to provide solutions to various global issues including international remittance for migrants. However, we have requirements for blockchain technology. To use blockchain and cryptoassets for international remittances, the price of the cryptoassets must be stable, the amount of energy required for transactions must be appropriate, the cost of the transaction must be low, the speed of the transaction must be high, and the occurrence of anomalies events such as money laundering, arbitrage transaction, and fraud must be prevented.

\begin{figure}[tbh]
\centering 
\includegraphics[width=0.8\textwidth]{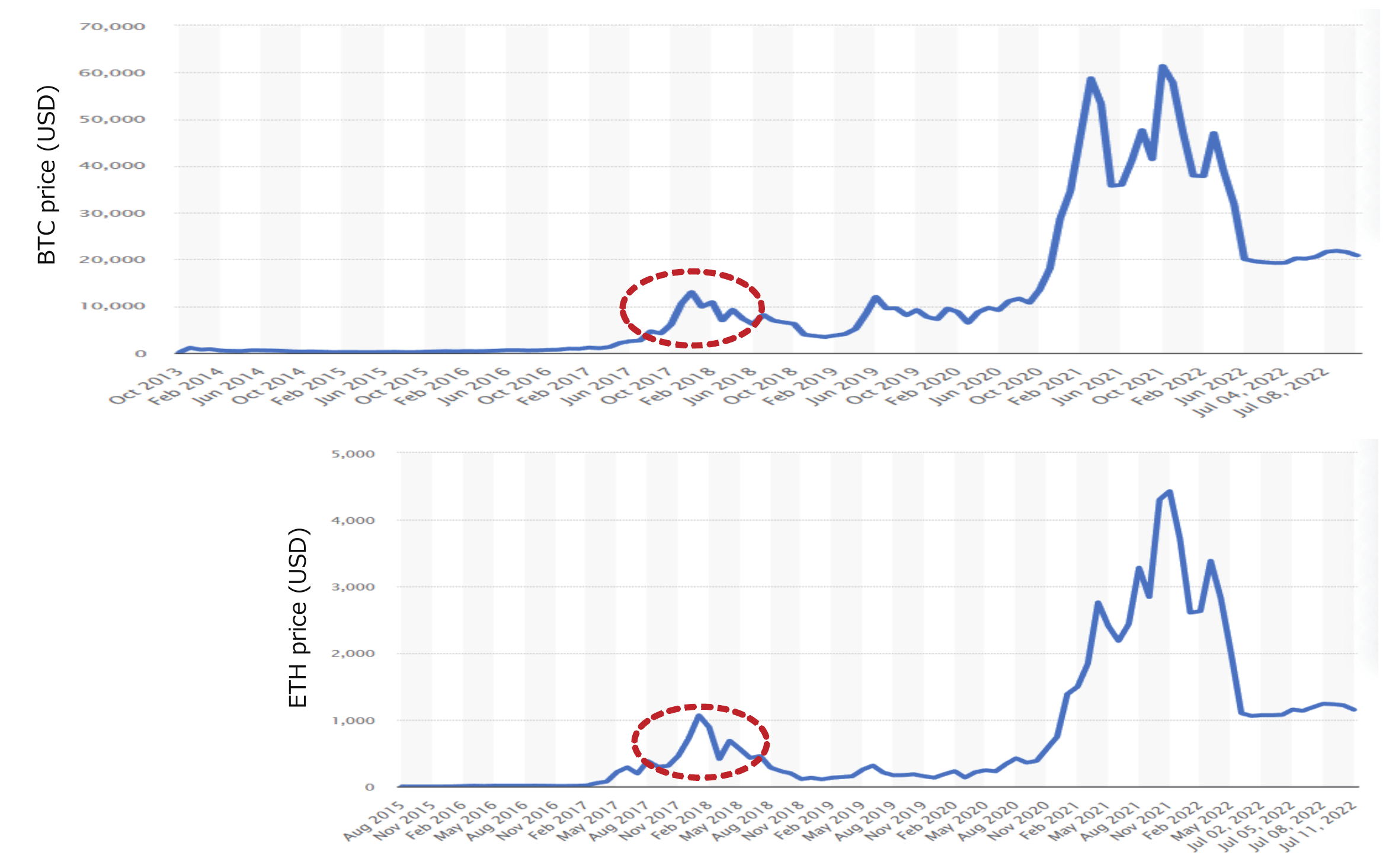}
\caption{\textbf{Temporal price changes of Bitcoin (top) and Ethereum (bottom).} Early 2018, with the price hike, is recognized as a bubble period.}
\label{f1}
\end{figure}

We attempt to detect anomalies in international remittances using cryptoassets during the bubble period of January 2018. 
This study analyzes the remittance transaction recorded on the XRP ledger for Ethereum (ETH) and United States dollar (USD) from July 2017 to Jun 2018, including the bubble period in early 2018 as shown in Fig. \ref{f1}. 
Figure \ref{f1} shows that Bitcoin (BTC) and ETH have a price hike from July 2017 to June 2018.
International remittance transaction is recorded as the amount of currency sent from sender (source) to recipient (destination) at time t. The remittance transactions made in a given period can be represented as a  remittance network.

The international remittance network is a directed graph consisting of a ``potential flow'' component and a ``loop flow'' component among sources and destinations.
The ``loop flow'' means that a remittance from one source (sender) goes through multiple destinations (recipients), some of which return to the original source.
In ordinary remittance transactions, the ``loop flow'' is unlikely to occur. Thus, the ``loop flow'' of remittances is considered an anomaly. 
In this paper, using the Hodge decomposition, we estimate the ``loop flow'' in the international remittance of cryptoassets during the bubble period to detect anomalies in international remittances. 
We present characteristic differences in the ``loop flow'' between fiat currencies and cryptoassets during the bubble period.

The remainder of the paper is organized as follows. We introduce Anomaly in Section \ref {Anomaly}. In Section \ref{Data}, we explain data recorded on XRP Ledger. In Section \ref{Analysis}, we discuss the basic methodology. Next, we discuss the results in Section \ref{Results and Discussions}. Finally, we provide some conclusions in Section \ref{Conclusions}.

\section{Anomaly Detection}
\label{Anomaly}

Blockchain technology's anonymity and anti-tamper are insufficient, and transactions of cryptoasset based on blockchain technologies may still be prone to various security, privacy, and reliability issues.
Chandola et al. \cite{Chandola2009} provide a structured and comprehensive overview of the research on anomaly detection. 
The authors have categorized anomalies: as point, contextual, and collective. 
For instance, fraud is categorized as a point anomaly.
The authors took credit card fraud, mobile phone fraud, insurance claim fraud, and insider trading in the stock market as examples of fraud.
For these fraud detections, statistical profiling using histograms and information-theoretic methodologies have proven to be very effective.
These studies of fraud detection for financial market transactions may provide a good reference for fraud detection for cryptoassets.

Hassan et al.~\cite{Hassan2022} surveyed the integration of anomaly detection models in blockchain technology.
The authors provided a detailed survey of the anomaly detection model in blockchain technology. They first discussed how anomaly detection could help secure blockchain-based applications. They then presented specific basic metrics and key requirements that can play an important role in developing anomaly detection for blockchain. They then surveyed the various anomaly detection models in terms of each layer of the blockchain (data, network, incentive, and smart contract layer). They found that (i) clustering is the most effective model for the data layer, (ii) general machine learning is the most effective model for the network layer, (iii) graph analysis is the most effective model for the incentive layer, and (iv) many models are used for detection in the smart contract layer.

In addition to these surveys, anomaly detection studies were conducted on trading networks of well-known cryptoassets. The following are examples of studies on Ethereum, bitcoin, and XRP.
Li et al.~\cite{Li2020} used topological data analysis to study the local topology of the Ethereum transaction network.
The local topology of the Ethereum transaction network contains a wealth of information on the crypto market, e.g., price prediction and price anomalies.
The authors proposed new functional summaries of topological descriptors, namely, Betti limits and Betti pivots, and showed that Betti pivots of the Ethereum transaction network improved precision in price anomaly prediction.

Hu et al.~\cite{Hu2019} studied the potential money laundering activities occurring in the Bitcoin network.
The authors found that the main difference between laundering and regular transactions lies in their output values and neighborhood information in the Bitcoin network. 
They evaluated four graph features: immediate neighbors, curated features, deepwalk embeddings, and node2vec embeddings to classify money laundering and regular transactions. 
The node2vec-based classifier outperforms other classifiers in binary classification.

Ikeda~\cite{Ikeda2022} studied the monthly direct transaction networks for cryptoasset XRP from the transactions recorded in XRP Ledger from January 2013 to September 2019. 
The author chose the 300 nodes from the entire network with the highest transaction value and built a network of all the nodes that have transactions with these 300 nodes and their 300 nodes to count the number of the 16 triangular motifs. 
Network motifs are small topological patterns, such as triangular sub-graphs, that recur in a network significantly more often than expected by chance. The statistically significant triangular motifs were identified by comparing the observed ratio with the theoretical expectation. These statistically significant motifs were more prevalent during the bubble-forming periods of 2014 and 2018, and less prevalent throughout the rest of the year. The result implies that the observed significant price change is caused by transactions corresponding to statistically significant triangular motifs, and thus detecting anomalies is possible by observing the triangular motifs.

\section{Data}
\label{Data}
The XRP ledger records two different types of data. One is a direct XRP Transaction, and the other is Settlement Transactions that transfer any type of credit, e.g., fiat currencies and cryptoassets.

\subsection{Direct XRP Transaction}
Individual users own their wallets on the XRP ledger. Different wallets may belong to the same users.
A type of credit, e.g., fiat currencies such as USD, EUR, and JPY, and cryptoassets such as XRP and BTC, is specified for a wallet.
Here we note that XRP is a cryptoasset that should be distinguished from the XRP ledger.
The hash public key identifies a wallet.
Direct XRP transactions allow the exchange of XRP between two wallets. This is the most usual form of XRP transaction.
For instance, user $u$ wants to pay $\beta$ XRP to user $v$, and $u$ has at least $\beta$ XRP in $u$’s XRP balance. Then $\beta$ XRP is removed from $u$’s XRP balance and added to $v$’s XRP balance on the XRP ledger.

\subsection{Settlement Transactions}
\begin{figure}[tbh]
\centering 
\includegraphics[width=0.5\textwidth]{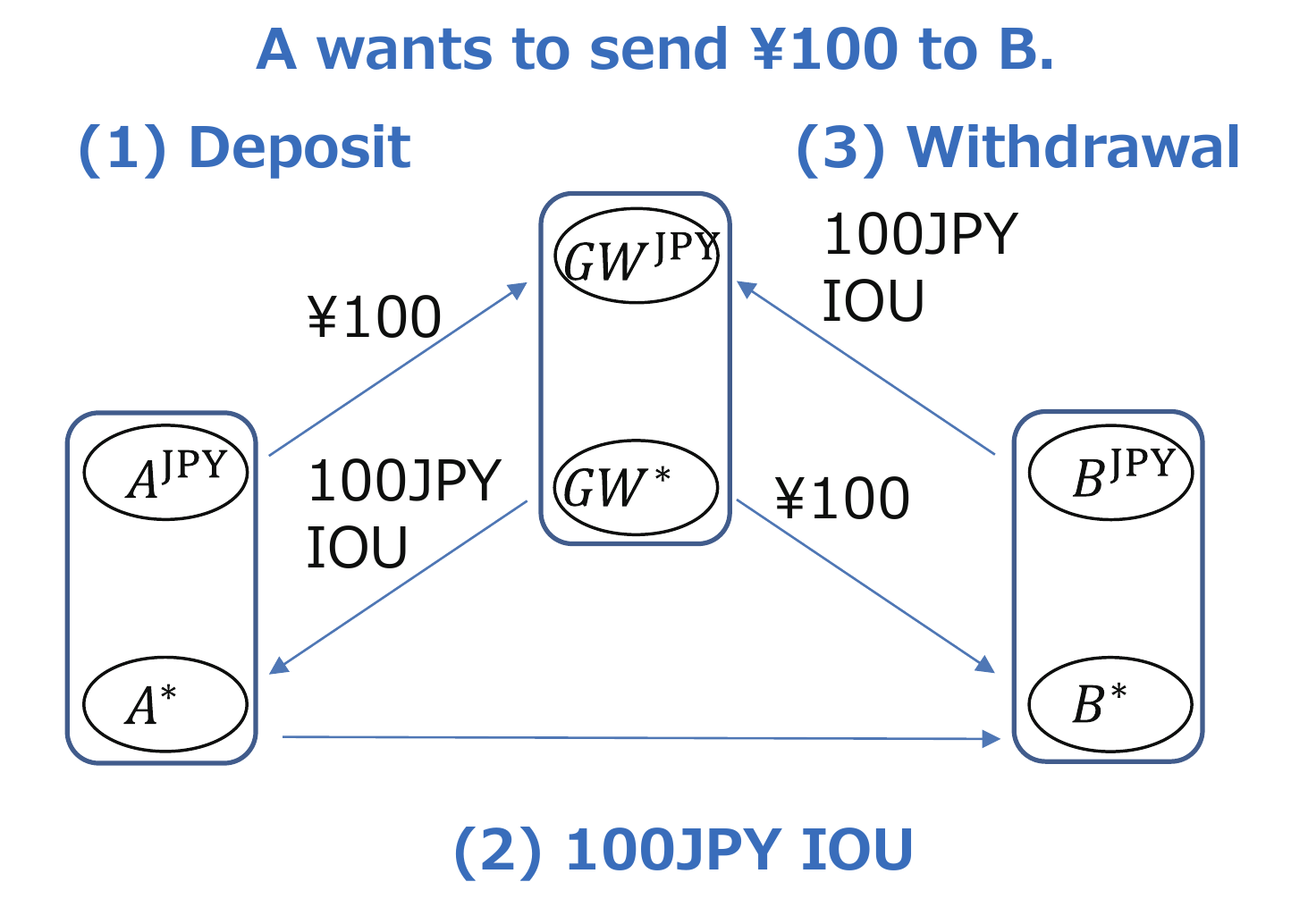}
\caption{\textbf{Remittance using Settlement Transactions} of $Y\llap{=} 100$ on the Ripple network from user $A$ to user $B$ via a gateway (GW). }
\label{f2}
\end{figure}

Settlement transactions transfer any type of credit (fiat currencies, cryptoassets, and user-defined currencies) between two wallets with a suitable set of credit paths on the Ripple network \cite{Moreno-Sanchez2016}.
On the Ripple network, remittance is done as settlement transactions that transfer a type of credit, e.g., fiat currencies and cryptoassets, between two wallets with suitable credit paths. 
The settlement transaction can only be performed by registered users on the Ripple network, but all transactions are recorded in the XRP Ledger and made available to the public.
Figure \ref{f2} explains a remittance transaction of $Y\llap{=} 100$ on the Ripple network from user $A$ to user $B$ via a gateway (GW).
User $A$ makes $Y\llap{=} 100$ deposit to GW, and GW issues an IOU to $A$ for $Y\llap{=} 100$. This IOU is sent from user $A$ to user $B$. User $B$ sends this IOU to GW and withdraws $Y\llap{=} 100$. At this time, GW's IOU disappears.
Gateway is a well-known reputed wallet on the Ripple Network that can trust to create and maintain an IOU credit correctly.
Here IOU credit guarantees a claim for the amount borrowed.
Gateway plays an essential role in remittance transactions on the Ripple Network. 
We note that IOU issuers are often used instead of a gateway.
IOU issuers can issue IOC credits, although they are less reputable wallets than gateways.
In this study, we do not distinguish between gateway and IOU issuer as having equivalent functions.

\begin{figure}[tbh]
\centering 
\includegraphics[width=0.8\textwidth]{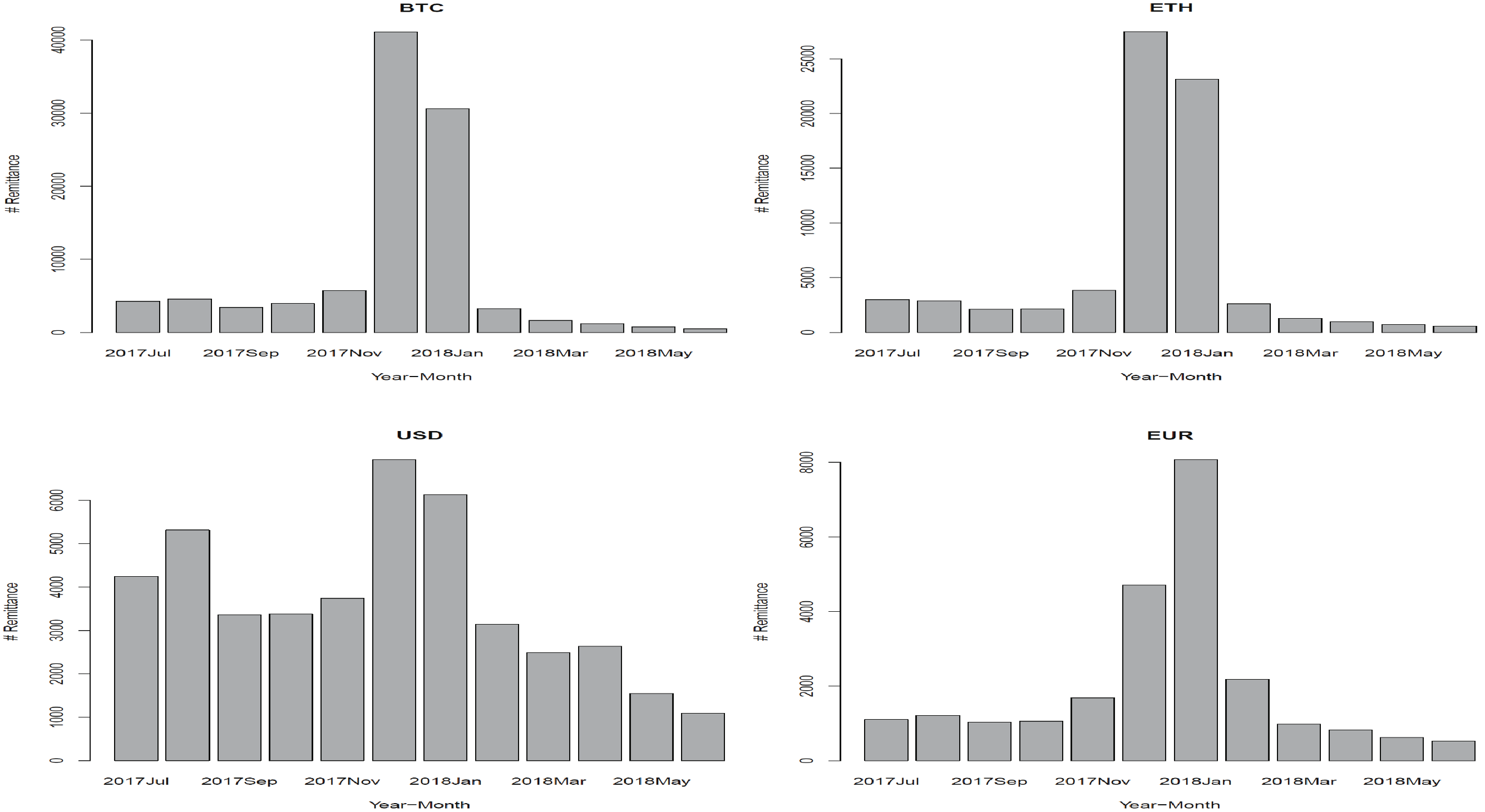}
\caption{\textbf{Monthly Change of the Number of Remittances for BTC, ETH, USD, and EUR.} In all assets and currencies, we observed significant increases in December 2017 and January 2018.}
\label{f3}
\end{figure}

The number of remittance transactions in each month from July 2017 to June 2018 is shown for BTC, ETH, USD, and EUR in Fig. \ref{f3}. In all assets and currencies, we observed significant increases in December 2017 and January 2018.

\section{Analysis Method}
\label{Analysis}

\subsection{Hypergraph}

\begin{figure}[tbh]
\begin{tabular}{cc}
\begin{minipage}[t]{0.45\hsize}
\centering
\includegraphics[width=0.7\textwidth]{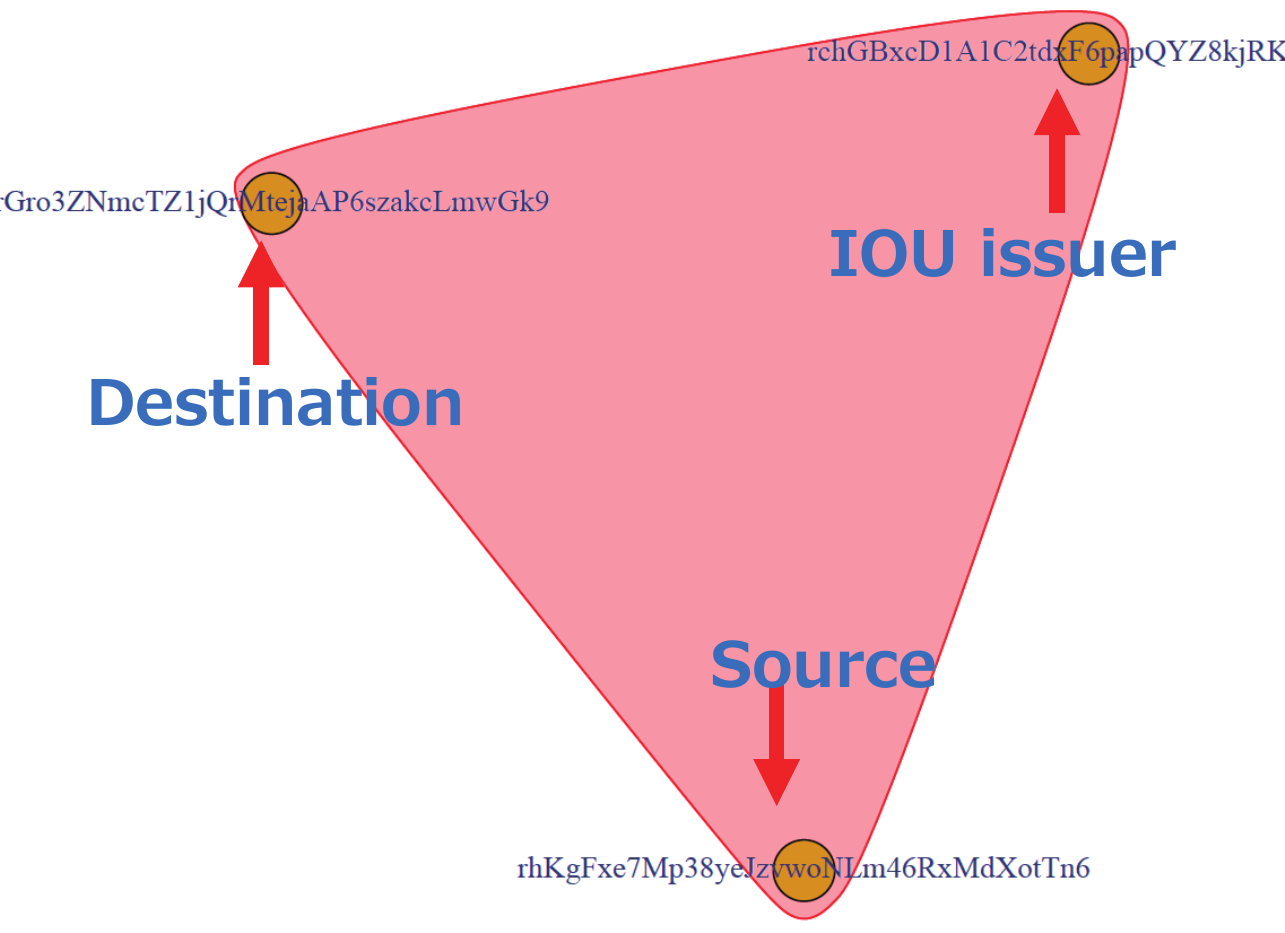}
\end{minipage} &
\begin{minipage}[t]{0.45\hsize}
\centering
\includegraphics[width=0.7\textwidth]{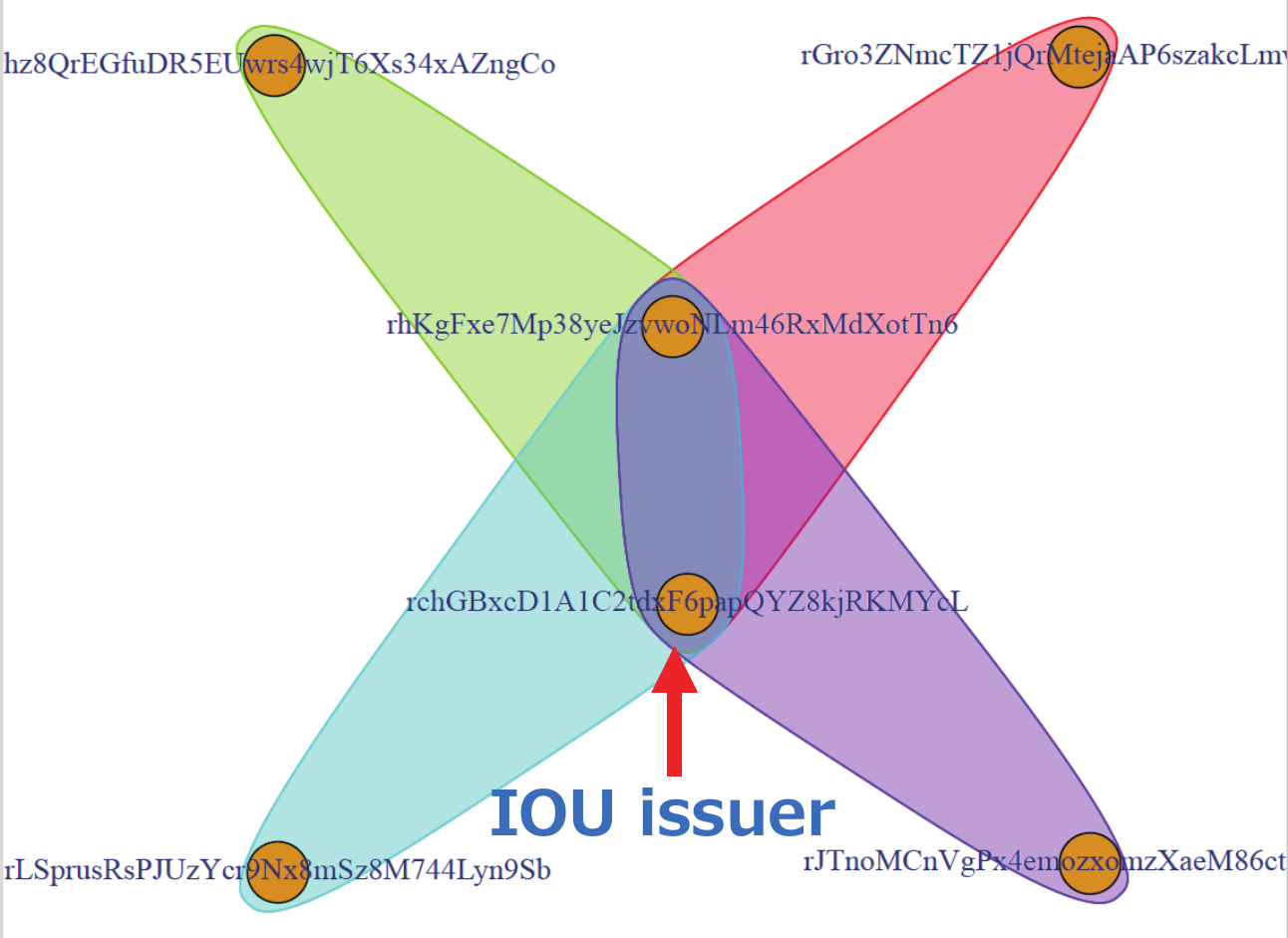}
\end{minipage}
\end{tabular}
\caption{\textbf{Remittance Recorded on the XRP Ledger.} The relationship, including more than two nodes, is called a higher-order interaction, and a network consisting of higher-order interactions is called a hypergraph.}
\label{f4}
\end{figure}

%

\begin{figure}[tbh]
\begin{tabular}{cc}
\begin{minipage}[t]{0.45\hsize}
\centering
\includegraphics[width=0.65\textwidth]{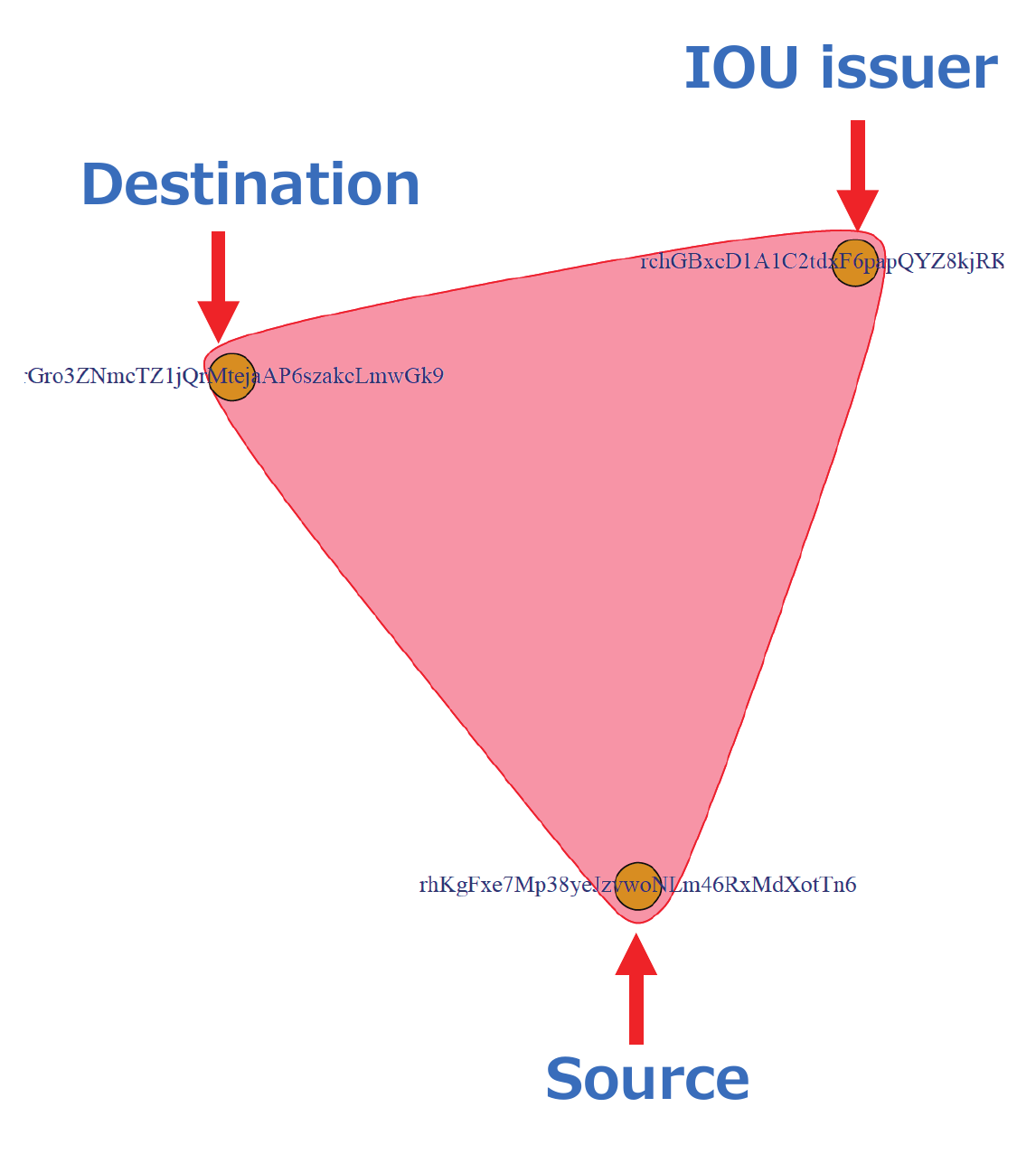}
\end{minipage} &
\begin{minipage}[t]{0.45\hsize}
\centering
\includegraphics[width=0.65\textwidth]{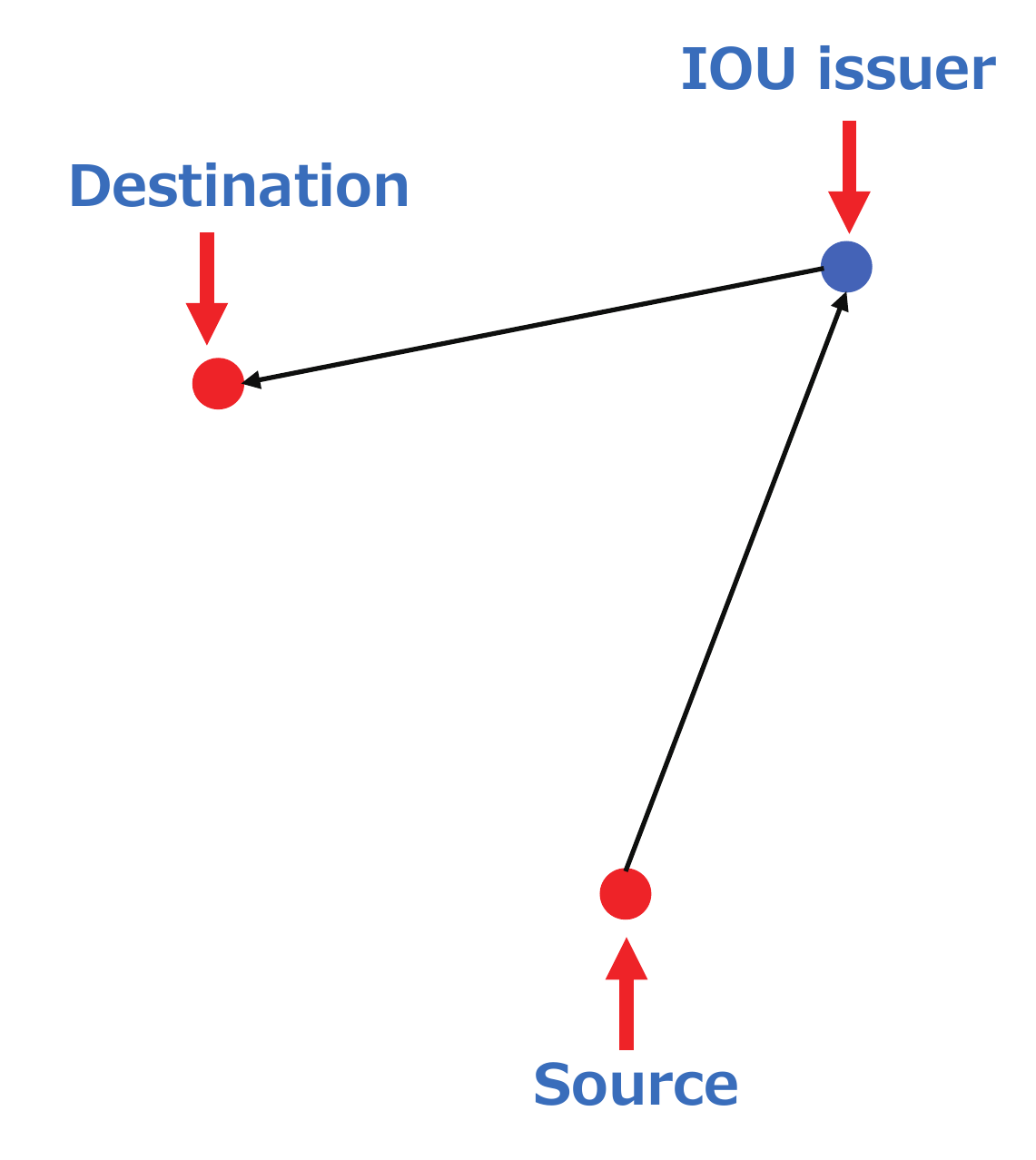}
\end{minipage}
\end{tabular}
\caption{\textbf{Disintegration of Higher-Order Interaction.} The three-body remittance interaction (left) disintegrates into a weighted directed link from a source to an IOU issuer and a subsequent weighted directed link from an IOU issuer to a destination (right).}
\label{f5}
\end{figure}


Remittance transactions are recorded on the XRP ledger.
Each remittance includes three nodes: a source, a destination, and an IOU issuer, as shown in the left panel of Fig. \ref{f4}. 
The hash public key is overlaid on top of each node.
The red-colored area surrounding the three nodes represents the remittance transaction relationship.
The right panel of Fig. \ref{f4} includes four remittance transactions. The source at the upper center sent some assets to four destinations at the four corners via an IOC issuer at the lower center. In this case, we have four transaction relationships in different colored areas.
The relationship, including more than two nodes, is called a higher-order interaction or hyperlink.
The remittance relationship can be represented by a hyperlink containing three nodes: a source, a destination, and an IOU issuer.
A network consisting of higher-order interactions or hyperlinks is called a higher-order network or a hypergraph \cite{Battiston2021, Bianconi2021}.
A hypergraph is a generalization of a graph consisting of edges (links) connecting two nodes, where an edge can connect any number of nodes. It is a method suitable for describing dynamics, such as synchronous and infectious phenomena in complex systems, and relationships between two or more parties, such as co-authorship in a paper.

Remittance using settlement transactions from a source to a destination is done via an IOU issuer, as shown in Fig. \ref{f2}.
Therefore, we approximate a hypergraph to a weighted directed graph without losing generality.
We disintegrate the hyperlink representing the three-body remittance interaction into the two subsequent links representing the two subsequent two-body interactions.
Figure \ref{f5} the three-body remittance interaction (left) disintegrates into a weighted directed link from a source to an IOU issuer and a subsequent weighted directed link from an IOU issuer to a destination (right).

\subsection{Hodge Decomposition}

We explain the Hodge decomposition \cite{Kichikawa2019, Fujiwara2020} to estimate the ``potential flow'' and ``loop flow'' in the international remittance of cryptoassets during the bubble period.
The higher-order interaction in the remittance network disintegrates into the two subsequent two-body interactions. In this approximation, we obtained a weighted directed network for international remittance.
A weighted directed network consisting of $N$ nodes is defined by adjacency matrix $A$ and weighted adjacency matrix $B$:
\begin{equation}
A = [ a_{ij} ]=
\left\{ 
\begin{array}{l}
1 \;\;\; (\text{if directed edge from $i$ to $j$}) \\
0 \;\;\; (\text{otherwise}),
\end{array} \right.
\label{adj}
\end{equation}
\begin{equation}
B = [ b_{ij} ]=
\left\{ 
\begin{array}{l}
b_{ij} \;\;\; (\text{if directed edge from $i$ to $j$ has a weight}) \\
0 \;\;\;\;\; (\text{otherwise}),
\end{array} \right.
\label{wghtadj}
\end{equation}
where $i=1,\cdots,N$ and $j=1,\cdots,N$. 

We define total flow matrix $F$ and weight matrix $W$ using  adjacency matrix $A$ and weighted adjacency matrix $B$ as follows:
\begin{equation}
F_{ij} = B_{ij} - B_{ji},
\label{flw}
\end{equation}
\begin{equation}
W_{ij} = A_{ij} + A_{ji}.
\label{wght}
\end{equation}
Graph Laplacian $L$ is written as follows using weight matrix $W$:
\begin{equation}
L_{ij} = \delta_{ij} \left( \sum_k W_{ik} \right) - W_{ij}.
\label{grlp}
\end{equation}
We obtain flow potential $\phi$ by solving the following Laplace-like equation numerically on the network:
\begin{equation}
\sum_j L_{ij} \cdot \phi_j = \sum_j F_{ij}.
\label{lpeq}
\end{equation}
Flow potential $\phi$ is arbitrary for the shift from the origin. We shift the origin to get $\sum_j \phi_j = 0$.
By taking numerical derivative of potential $\phi$, we obtain potential flow matrix $F^{pot}$:
\begin{equation}
F_{ij}^{pot} = W_{ij} \cdot  \left( \phi_i - \phi_j \right). 
\label{hodge1}
\end{equation}
Finally, we obtain loop flow matrix $F^{loop}$ by subtracting $F^{pot}$ from total flow matrix $F$ as follows:
\begin{equation}
F_{ij}^{loop} = F_{ij} - F_{ij}^{pot} = F_{ij} - W_{ij} \cdot  \left( \phi_i - \phi_j \right).
\label{hodge2}
\end{equation}
This procedure is called Hodge decomposition on a weighted directed network.

\section{Results and Discussions}
\label{Results and Discussions}
We show that results of the network analysis for the remittance transaction recorded on the XRP ledger for ETH and USD from July 2017 to Jun 2018, including the bubble period in early 2018.

\subsection{ETH Remittance}
\begin{figure}[tbh]
\centering 
\includegraphics[width=0.3\textwidth]{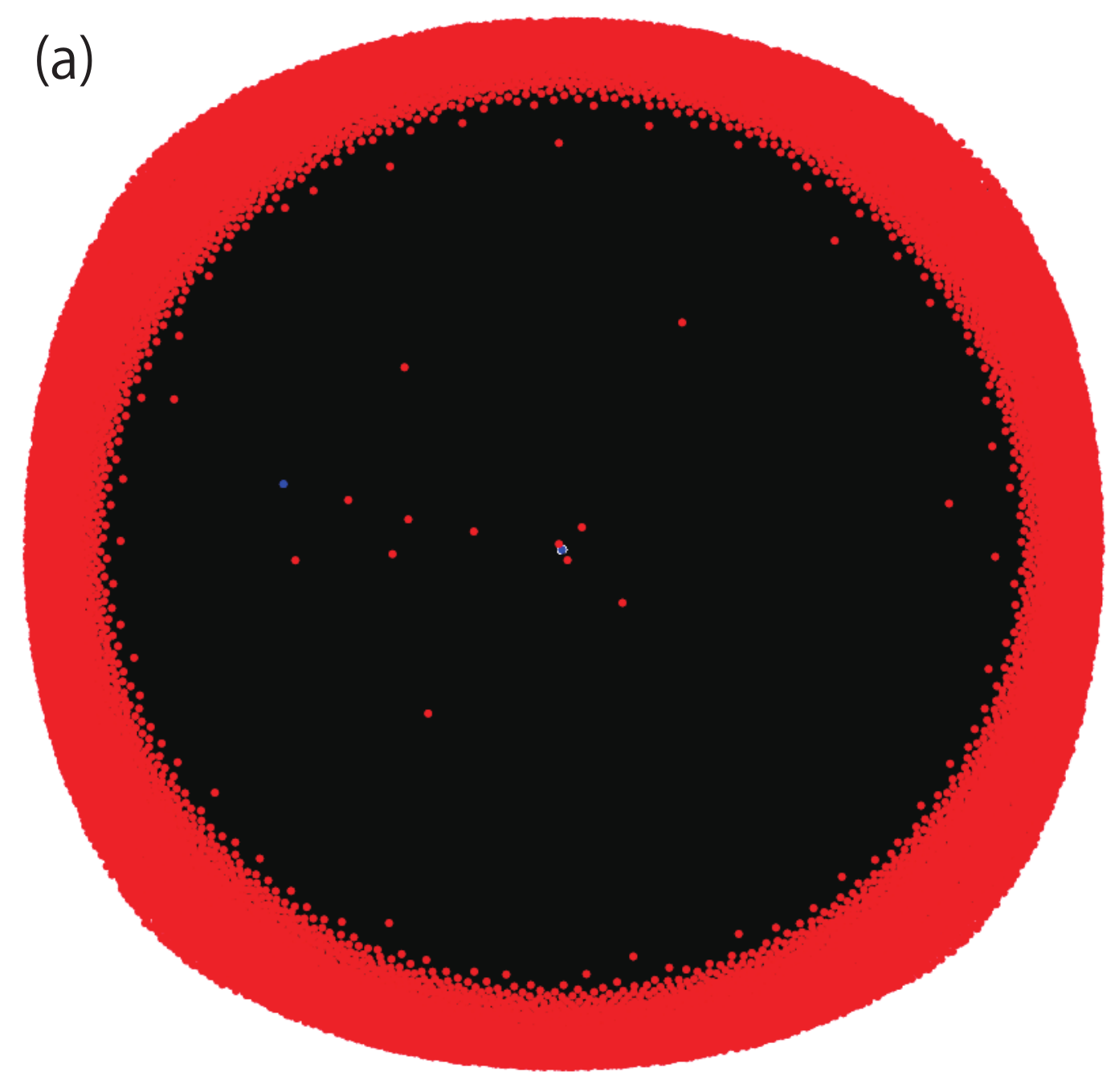}
\includegraphics[width=0.69\textwidth]{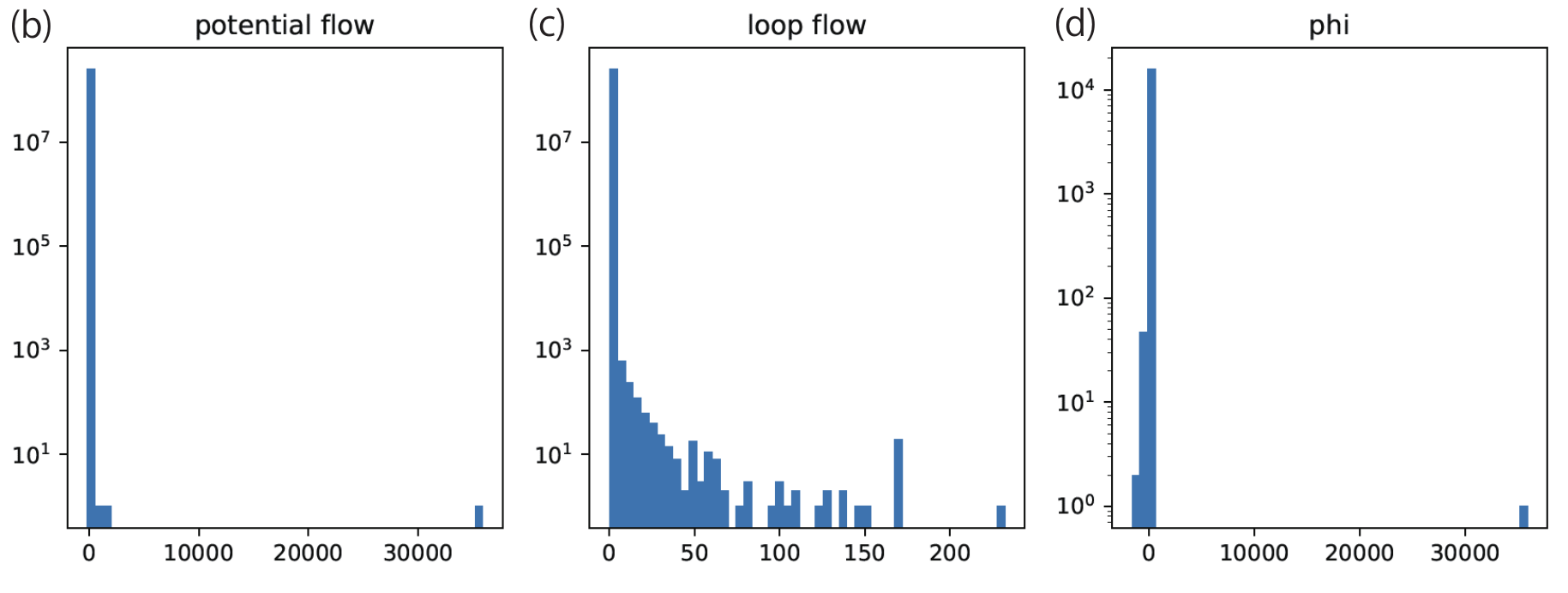}
\caption{\textbf{ETH Remittance in December 2017.} (a) Remittance network diagram, (b) Distribution of $F_{ij}^{pot}$, (c) Distribution of $F_{ij}^{loop}$, and (d) Distribution of $\phi_i$}
\label{f6}
\end{figure}

\begin{figure}[tbh]
\centering 
\includegraphics[width=0.3\textwidth]{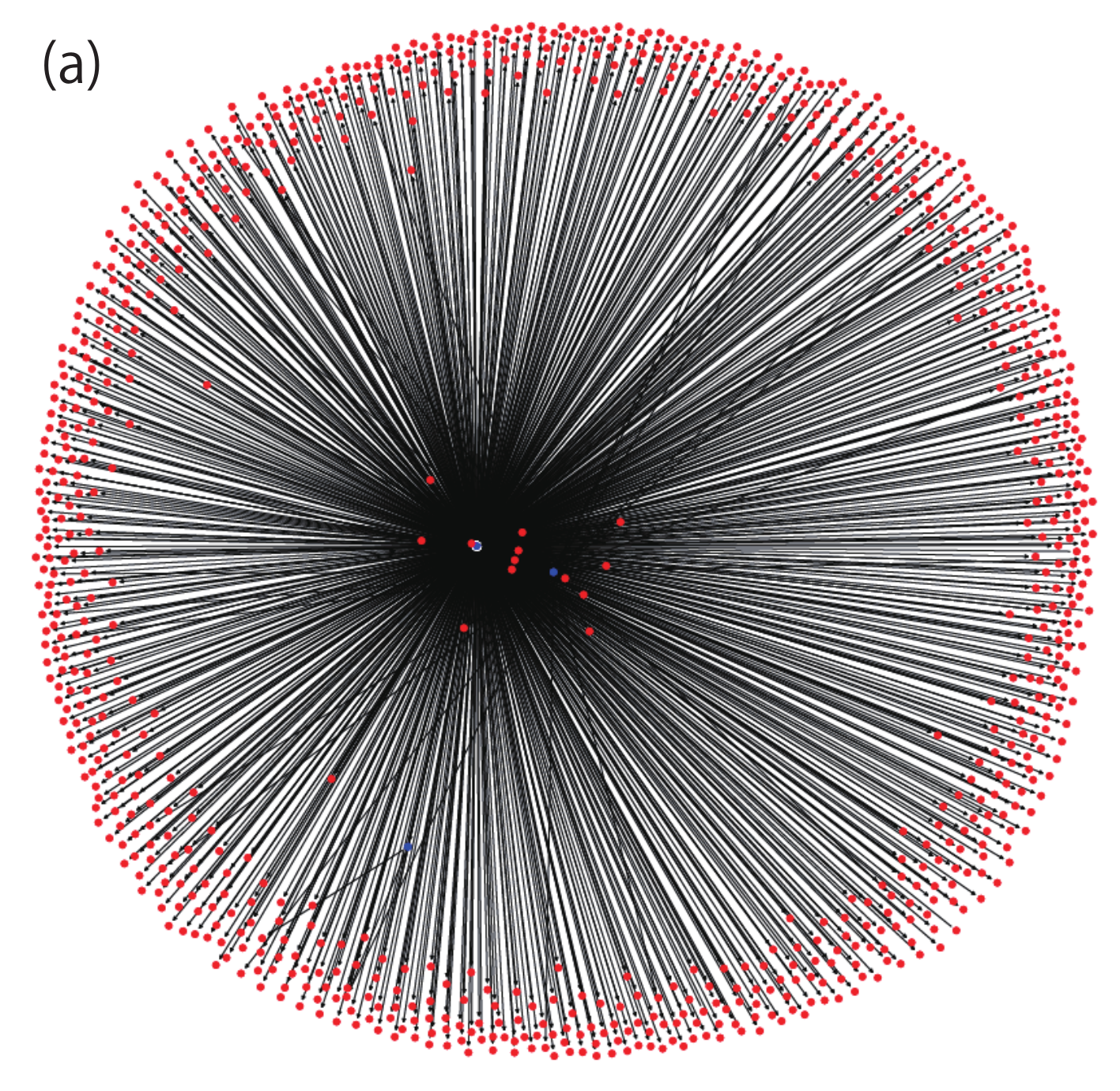}
\includegraphics[width=0.69\textwidth]{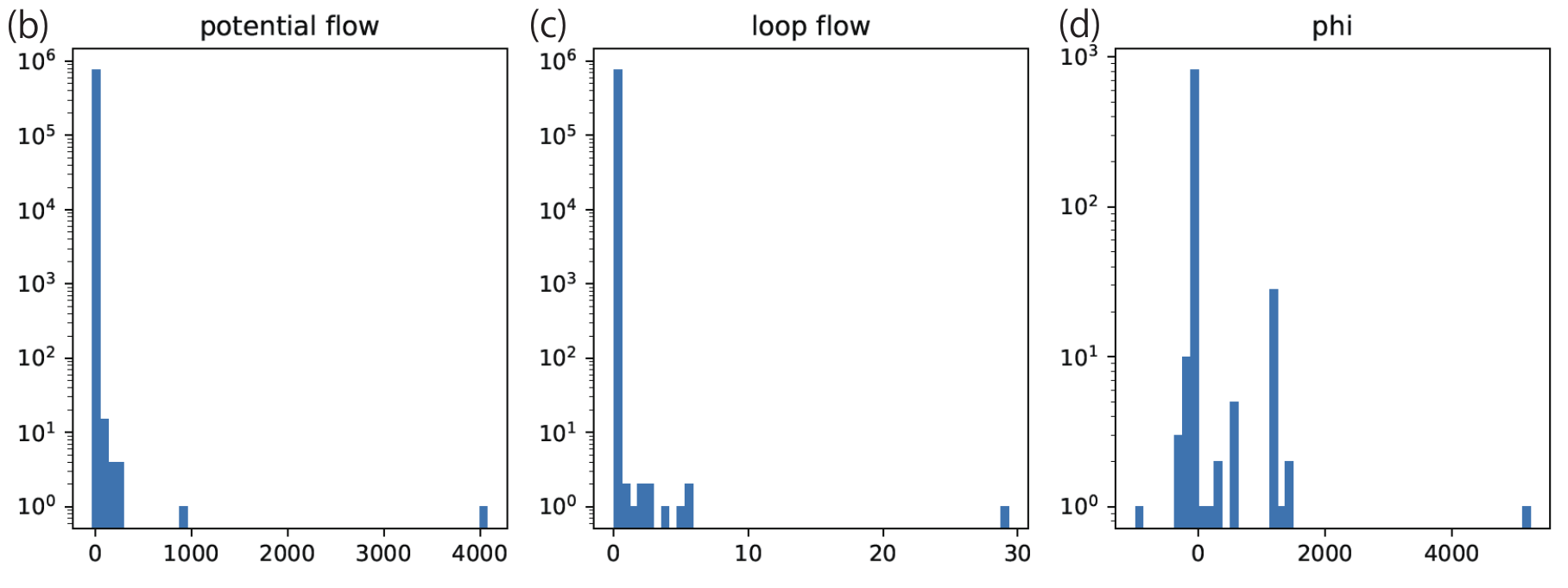}
\caption{\textbf{ETH Remittance in May 2018.} (a) Remittance network diagram, (b) Distribution of $F_{ij}^{pot}$, (c) Distribution of $F_{ij}^{loop}$, and (d) Distribution of $\phi_i$}
\label{f7}
\end{figure}

Panel (a) of Figs. \ref{f6} and \ref{f7} show remittance network diagrams in December 2017 and May 2018, respectively.
Nodes are indicated by red circles and linked by black lines with arrows. Sources are located in the center of the network, and destinations are located on the periphery; IOU issuers are indicated by blue circles: two IOU issuers in December 2017 and three IOU issuers in January 2018.
For ETH remittance, a comparison between a panel (c) of Figs. \ref{f6} and \ref{f7} shows that 
there was a significant increase in the loop-flow component during the cryptoasset price peak in December 2017.

\begin{figure}[tbh]
\centering 
\includegraphics[width=0.8\textwidth]{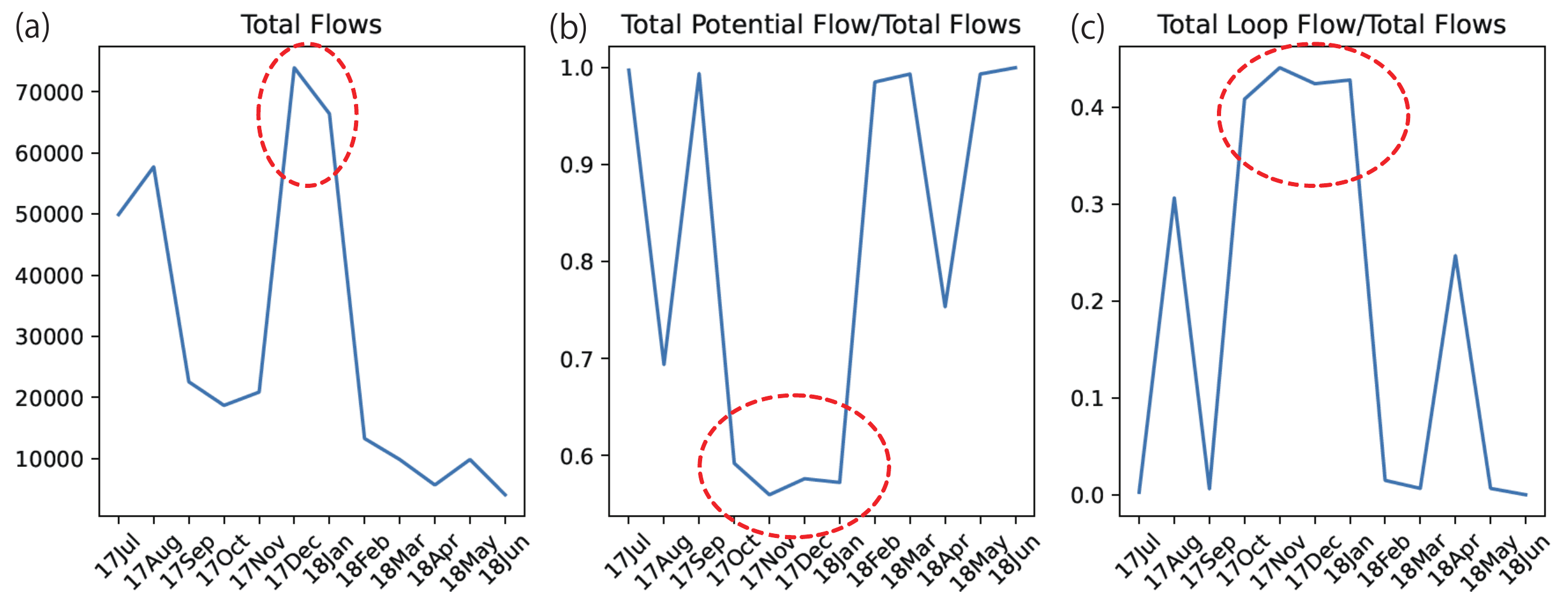}
\caption{\textbf{Monthly Change of Potential and Loop Flow of Remittance for ETH.} (a) Total flow $F^{total}=\sum_{ij}F_{ij}$, (b) Ratio of potential flow to total flow $\sum_{ij}F_{ij}^{pot}/F^{total}$, (c) Ratio of loop flow to total flow $\sum_{ij}F_{ij}^{loop}/F^{total}$}
\label{f8}
\end{figure}

The monthly change of potential and loop flow of remittance for ETH are shown in Fig. \ref{f8}.
Panel (a) shows total flow $F_{toatl}=\sum_{ij}F_{ij}$. We observed a prominent increase in remittances between December 2017 and January 2018.
Panel (b) shows $\sum_{ij}F_{ij}^{pot}/F_{toatl}$ and panel (c) shows $\sum_{ij}F_{ij}^{loop}/F_{toatl}$.
We observed in these two panels a prominent increase in ``loop flow'' up to approximately $40\%$ of total flow and a decrease in ``potential flow'' when the price hiked in December 2017 and January 2018.
As discussed earlier, the ``loop flow'' is unlikely to occur in ordinary remittance transactions. For this reason, we consider remittances' ``loop flow'' as an anomaly. 
The observed increase in ``loop flow'' at the price hike might be caused by money laundering or arbitrage transactions.
If the geographical location of the largest IOU issuer is identified, it would be helpful to determine the cause of the increased loop flow.

%


\subsection{USD Remittance}

\begin{figure}[tbh]
\centering 
\includegraphics[width=0.3\textwidth]{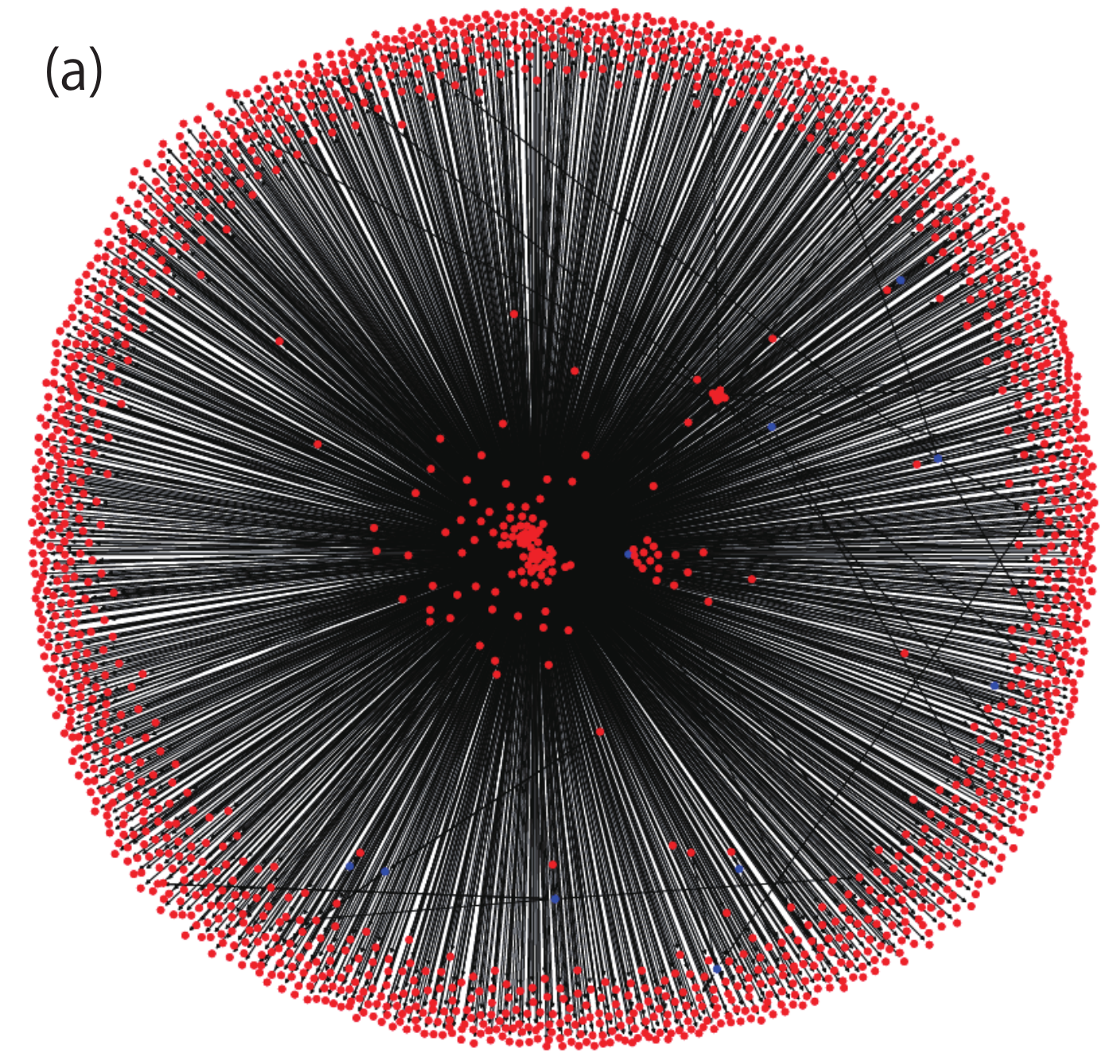}
\includegraphics[width=0.69\textwidth]{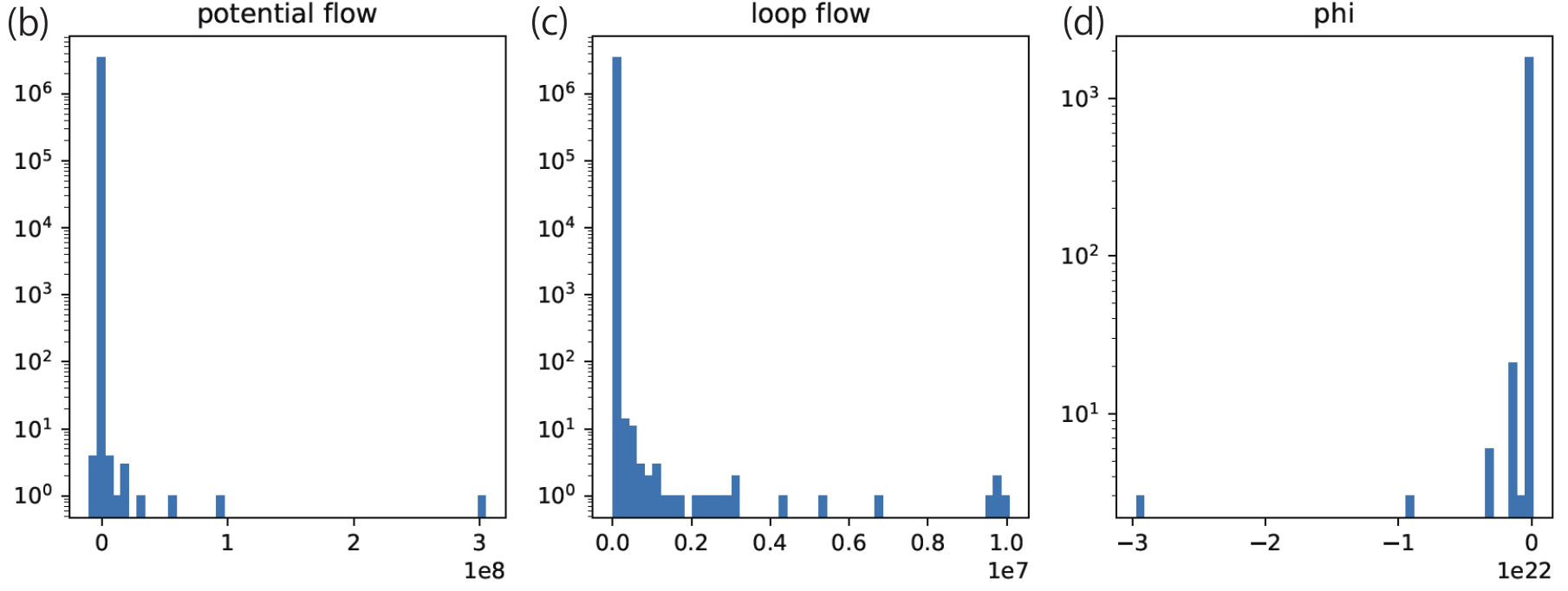}
\caption{\textbf{USD Remittance in January 2018.} (a) Remittance network diagram, (b) Distribution of $F_{ij}^{pot}$, (c) Distribution of $F_{ij}^{loop}$, and (d) Distribution of $\phi_i$}
\label{f10}
\end{figure}

\begin{figure}[tbh]
\centering 
\includegraphics[width=0.3\textwidth]{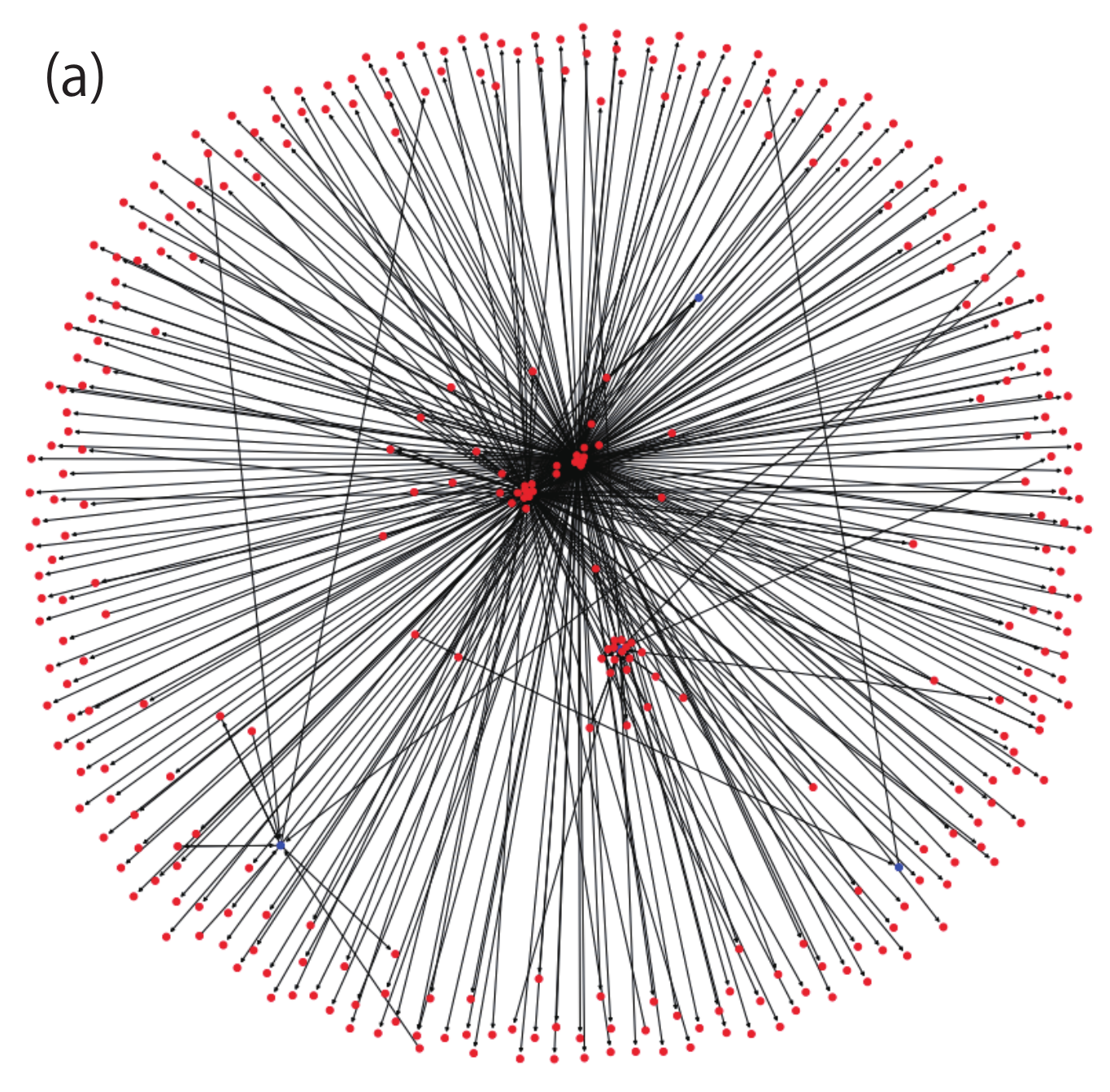}
\includegraphics[width=0.69\textwidth]{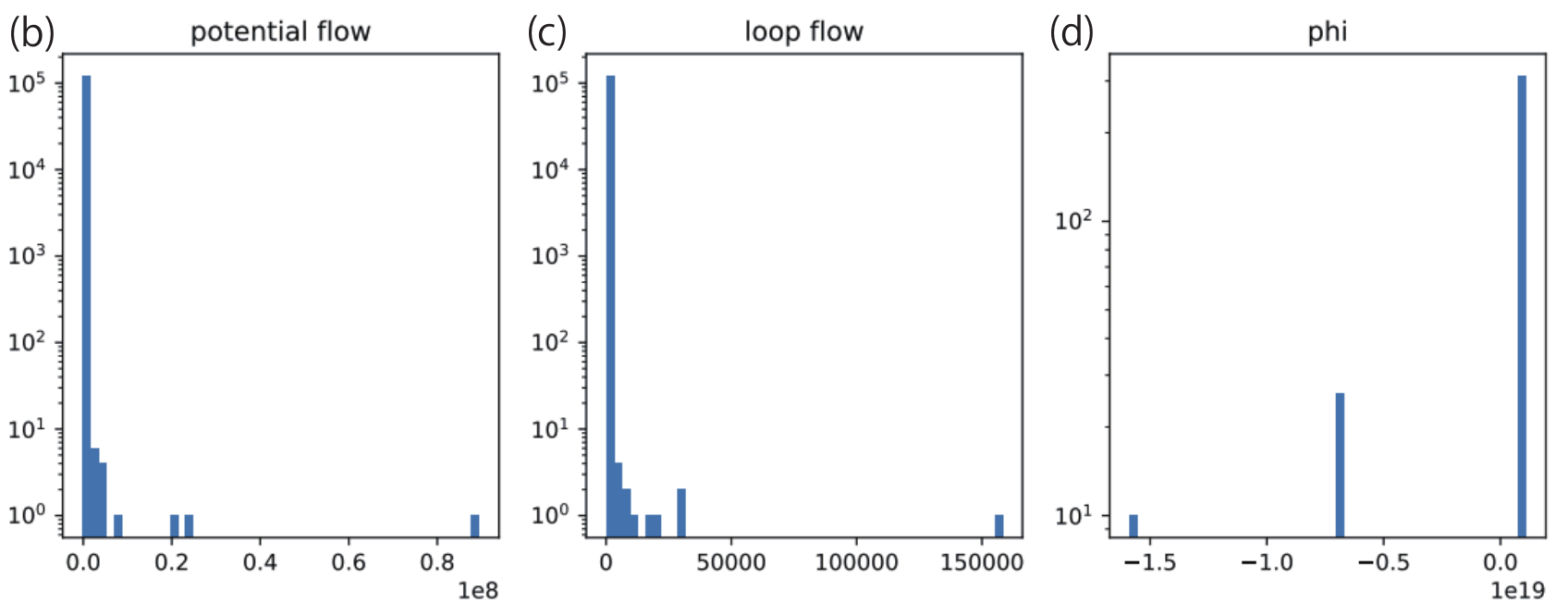}
\caption{\textbf{USD Remittance in April 2018.} (a) Remittance network diagram, (b) Distribution of $F_{ij}^{pot}$, (c) Distribution of $F_{ij}^{loop}$, and (d) Distribution of $\phi_i$}
\label{f11}
\end{figure}

Panel (a) of Figs. \ref{f10} and \ref{f11} show remittance network diagrams in January 2018 and April 2018, respectively.
Nodes are indicated by red circles and linked by black lines with arrows. Sources are located in the center of the network, and destinations are located on the periphery; IOU issuers are indicated by blue circles.
For USD remittance, a comparison between a panel (c) of Figs. \ref{f10} and \ref{f11} shows that 
there was a slight increase in the loop-flow component for USD remittance during the cryptoasset price peak in December 2017.

\begin{figure}[tbh]
\centering 
\includegraphics[width=0.8\textwidth]{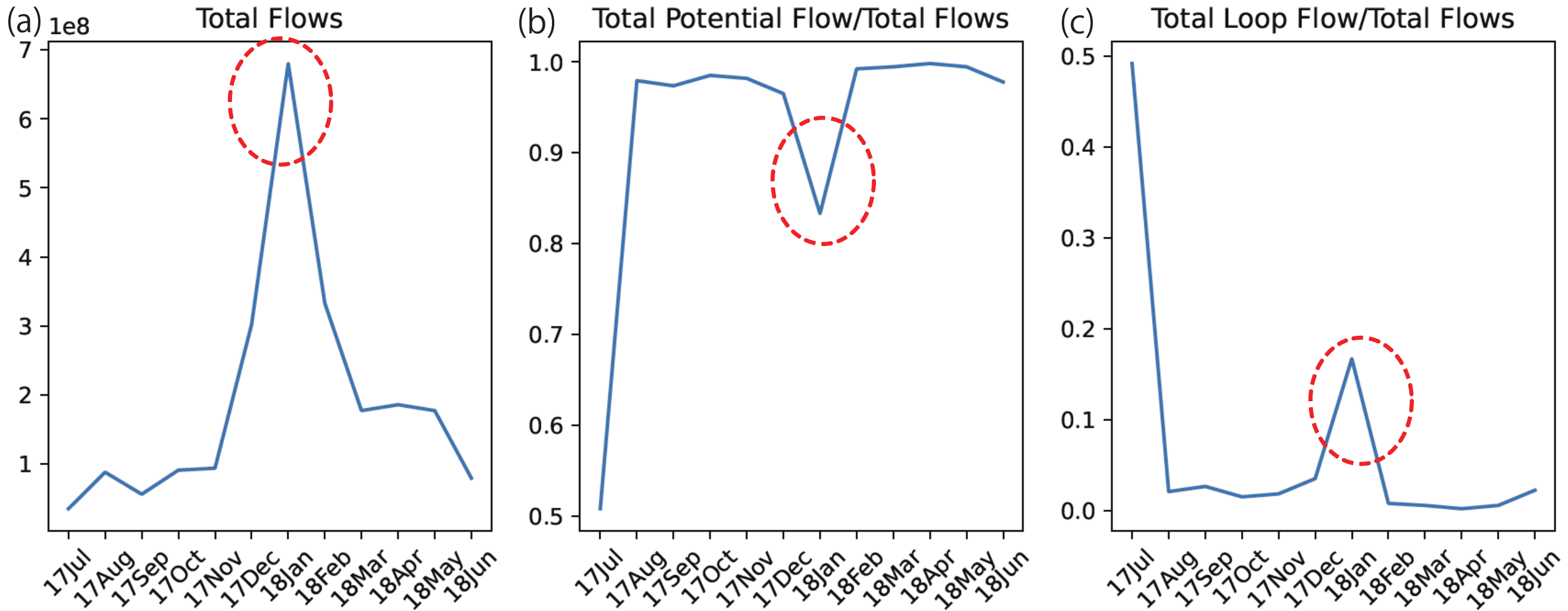}
\caption{\textbf{Monthly Change of Potential and Loop Flow of Remittance for USD.} (a) Total flow $F^{total}=\sum_{ij}F_{ij}$, (b) Ratio of potential flow to total flow $\sum_{ij}F_{ij}^{pot}/F^{total}$, (c) Ratio of loop flow to total flow $\sum_{ij}F_{ij}^{loop}/F^{total}$}
\label{f12}
\end{figure}

The monthly change of potential and loop flow of remittance for USD is shown in Fig. \ref{f12}.
Panel (a) shows total flow $F^{total}=\sum_{ij}F_{ij}$. We observed a significant increase in remittances between December 2017 and January 2018.
Panel (b) shows $\sum_{ij}F_{ij}^{pot}/F^{total}$ and panel (c) shows $\sum_{ij}F_{ij}^{loop}/F^{total}$.
We observed in these two panels a small increase in ``loop flow'' and a small decrease in ``potential flow'' when the price hiked in December 2017 and January 2018.
Cryptoassets are traded on various markets, but prices vary slightly from market to market. On the other hand, Fiat currency has a constant exchange rate regardless of the market. Therefore, arbitrage is likely to occur in cryptoassets.

%

\section{Conclusions}
\label{Conclusions}
We analyzed the remittance transaction recorded on the XRP ledger for BTC and ETH from July 2017 to Jun 2018, including the bubble period in early 2018. 
We attempt to detect anomalies in international remittances using cryptoassets during the bubble period of January 2018. In ordinary remittance transactions, the ``loop flow'' is unlikely to occur. Thus, the ``loop flow'' of remittances is considered an anomaly. 
Using the Hodge decomposition, we estimate the ``loop flow'' in the international remittance of cryptoassets during the bubble period to detect anomalies in international remittances. 
We present characteristic differences in the ``loop flow'' between fiat currencies and cryptoassets during the bubble period.
We found characteristic differences between cryptoasset ETH and fiat currency USD during the bubble period.
For ETH, there was a significant increase in the loop flow during the cryptoasset price peak. This might be related to money laundering or arbitrage transaction.
There was a slight increase in the loop flow for USD during the cryptoasset price peak.
Cryptoassets are traded on various markets, but prices vary slightly from market to market. On the other hand, Fiat currency has a constant exchange rate regardless of the market. Therefore, arbitrage is likely to occur in cryptoassets.


\section*{Acknowledgement}
Author (YI) thanks Yoshi Fujiwara, Hiroshi Iyetomi, and Yuji Hirono for their helpful discussion and comments on the formulation of the Hodge decomposition.
This work is partially supported by Ripple Impact Fund 2022-247584 (5855).


\end{document}